\documentclass[fleqn,usenatbib]{mnras}

\usepackage{graphicx}
\usepackage{amsmath,amssymb}
\usepackage{pdflscape}
\usepackage{newtxtext,newtxmath}

\usepackage[T1]{fontenc}

\DeclareRobustCommand{\VAN}[3]{#2}
\let\VANthebibliography\thebibliography
\def\thebibliography{\DeclareRobustCommand{\VAN}[3]{##3}\VANthebibliography}

\title[Protocluster at $z=3.7$]{Galaxy properties from the outskirts to the core of a protocluster at $\mathbf{z=3.699}$}

\author[J. Toshikawa et al.]{
Jun Toshikawa,$^1$\thanks{E-mail: toshikawa@nhao.jp}
Stijn Wuyts,$^2$ Nobunari Kashikawa,$^3$ Hisakazu Uchiyama,$^4$ Malcolm Bremer,$^5$
\newauthor Marcin Sawicki,$^6$ Yoshiaki Ono,$^7$ Mariko Kubo,$^8$ Kei Ito$^3$\\
$^1$Nishi-Harima Astronomical Observatory, Center for Astronomy, University of Hyogo, Sayo, Hyogo 679-5313, Japan\\
$^2$Department of Physics, University of Bath, Claverton Down, Bath, BA2 7AY, UK\\
$^3$Department of Astronomy, University of Tokyo, Hongo, Tokyo 113-0033, Japan\\
$^4$National Astronomical Observatory of Japan, 2-21-1 Osawa, Mitaka, Tokyo 181-8588, Japan\\
$^5$H H Wills Physics Laboratory, University of Bristol, Tyndall Avenue, Bristol, BS8 1TL, UK\\
$^6$Department of Astronomy \& Physics and Institute for Computational Astrophysics, Saint Mary's University, 923 Robie Street, Halifax, NS B3H 3C3, Canada\\
$^7$Institute for Cosmic Ray Research, The University of Tokyo, 5-1-5 Kashiwa-no-Ha, Kashiwa, Chiba 277- 8582, Japan\\
$^8$Astronomical Institute, Tohoku University, 6-3, Aramaki, Aoba, Sendai, Miyagi, 980-8578, Japan\\
}

\date{Accepted XXX. Received YYY; in original form ZZZ}

\pubyear{2024}

\begin{document}
\label{firstpage}
\pagerange{\pageref{firstpage}--\pageref{lastpage}}
\maketitle

\begin{abstract}
We present follow-up spectroscopy on a protocluster candidate selected from the wide-field imaging of the Hyper
SuprimeCam Subaru Strategic Programme.
The target protocluster candidate was identified as a $4.5\sigma$ overdense region of $g$-dropout galaxies, and
the redshifts of $g$-dropout galaxies are determined by detecting their Ly$\alpha$ emission.
Thirteen galaxies, at least, are found to be clustering in the narrow redshift range of $\Delta z<0.05$ at
$z=3.699$.
This is clear evidence of the presence of a protocluster in the target region.
Following the discovery of the protocluster at $z=3.699$, the physical properties and three-dimensional
distribution of its member galaxies are investigated.
Based on spectroscopically-confirmed $g$-dropout galaxies, we find an overabundance of rest-frame ultraviolet (UV)
bright galaxies in the protocluster.  
The UV brightest protocluster member turns out to be an active galactic nucleus, and the other UV brighter members 
tend to show smaller Ly$\alpha$ equivalent widths than field counterparts.
The member galaxies tend to densely populate near the centre of the protocluster, but the separation from the
nearest neighbour rather than the distance from the centre of the protocluster is more tightly correlated to
galaxy properties, implying that the protocluster is still in an early phase of cluster formation and only close
neighbours have a significant impact on the physical properties of protocluster members. 
The number density of massive galaxies, selected from an archival photometric-redshift catalogue, is higher near
the centre of the protocluster, while dusty starburst galaxies are distributed on the outskirts.
These observational results suggest that the protocluster consists of multiple galaxy populations, whose spatial
distributions may hint at the developmental phase of the galaxy cluster.
\end{abstract}

\begin{keywords}
galaxies: evolution -- galaxies: high-redshift
\end{keywords}

\section{Introduction}

Galaxy clusters are the most massive gravitationally-bound structures in the universe, formed from
small perturbations in the density field when the universe was young.
In the local universe, galaxy clusters occupy a unique position within the large-scale structure, namely at the
nodes of the cosmic web \citep[e.g.,][]{peebles80,alpaslan14,libeskind18}; thus, the abundance or spatial
distribution of galaxy clusters are powerful tools to constrain cosmological parameters
\citep[e.g.,][]{hoessel80,fumagalli24,ghirardini24}.
In addition, cluster galaxies possess distinguishing characteristics from field galaxies
\citep[e.g.,][]{dressler80,wetzel12,gallazzi21}, suggesting that galaxy clusters have played a key role as the
drivers of environmental effects on galaxy evolution.
Consequently, galaxy clusters are a crucial bridge between galaxy evolution and the growth of cosmic structures,
and equivalently between astrophysical and cosmological phenomena.

It remains challenging to predict how galaxies formed and evolved within a cosmological framework.
The combination of $N$-body dark matter simulations and semi-analytic galaxy formation models enables such
modelling over large cosmological volumes \citep[e.g.,][]{springel05}, though a significant number of parameters
need to be tuned to reproduce observational results of galaxy properties \citep[e.g.,][]{ayromlou19,henriques20}.
While hydrodynamical simulations can more directly trace the effects of physical processes
\citep[e.g.,][]{barnes17,nelson23}, their larger computational expense introduces an unavoidable trade-off between
mass resolution and size of the cosmological box being simulated.
Therefore, it is also imperative to -alongside modelling endeavours- directly observe the developmental phase of
galaxy clusters, so-called ``protoclusters.''

Theoretical models predict that the size of protoclusters, or the spatial distribution of member galaxies, ranges
from $\sim1\,\mathrm{proper\,Mpc}$ (pMpc) up to $\sim10\,\mathrm{pMpc}$ depending on their descendant halo mass at
$z=0$ \citep{chiang13,muldrew15}.
Furthermore, the shape of protoclusters is not spherical, but reflects the cosmic web \citep{lovell18}.
Even at $z\lesssim0.5$, about three quarters of galaxy clusters remain dynamically unrelaxed
\citep[e.g.,][]{wen13,yuan22}.

Observationally, protoclusters are identified by overdense regions of high-redshift galaxies, though many
(slightly) different selection techniques are applied.
The main challenge to protocluster searches is the rarity of protoclusters.
To overcome this, some studies utilise quasars (QSOs) or radio galaxies (RGs) as signposts, because such massive
galaxies are expected to reside in highly biased environments \citep[e.g.,][]{venemans07,wylezalek13,luo22}.
Similarly, Ly$\alpha$ blobs (LABs) and bright sub-mm galaxies (SMGs) have also been successfully targeted to find
protoclusters \citep[e.g.,][]{matsuda11,clements16,calvi23}.
These signpost techniques would be efficient, but the relation between protoclusters and such peculiar galaxies
(e.g., whether every protocluster hosts them) is still under debate \citep[e.g.,][]{hatch14,uchiyama18,gao22}.
An alternative approach consists of mapping  sufficiently wide areas on the sky, without requiring any extreme
signposts as a prior, thus enabling the construction of a less biased sample of protoclusters
\citep[e.g.,][]{ouchi05,toshikawa12,lemaux18}.

Making full use of various techniques, the number of known protoclusters are increasing
\citep{overzier16,alberts22}.
In contrast to mature clusters as seen in the local universe, protoclusters are composed of star-forming galaxies,
and the relation between star-formation rate (SFR) and density is reversed \citep{lemaux22,shi24}.
Protoclusters at high redshifts are expected to account for a significant fraction of cosmic SFR in spite of the
rarity of these compact structures \citep[e.g.,][]{casey16,kato16}.
Some extreme examples of protoclusters reach total star formation rates of
$\gtrsim10^4\,\mathrm{M_{\sun}\,yr^{-1}}$ in a single system \citep{miller18,oteo18,hill20}.
Such intense star-forming activity may be sustained by a large amount of gas accreting along the cosmic web   
\citep{umehata19,daddi22}.
An increased frequency of mergers/interactions may further boost the total SFR by triggering starburst phases.
On the other hand, massive, quiescent galaxies also exist in protoclusters, at least as early as
$z\sim2\mathrm{-}4$ \citep{kubo21,ito23,tanaka23}.
\citet{toshikawa16} confirmed several protoclusters from the same survey with a consistent method; however, they
exhibit different features from each other.
\citet{toshikawa20} discovered two protoclusters at $z\sim3.7$: one shows a centrally concentrated distribution
of protocluster members like a core, while the other is divided into several sub-groups. 
In addition, the fraction of massive quiescent galaxies among protocluster galaxies also differs from sample to
sample even at the same redshift \citep{shi20,shi21}.
As the history of cluster formation develops over Hubble timescales, protocluster structures and properties can
vary significantly depending on their developmental phase.
The descendant halo mass at $z = 0$ is an important parameter in this context, but in practice can only be
empirically constrained in a statistical manner \citep{press74}.
Even a precise measurement of the halo mass at the (high-redshift) epoch of observation will result in a
$\sim1\,\mathrm{dex}$ uncertainty on the descendant halo mass at $z=0$.
This large range makes it difficult to fairly compare protoclusters, further compounded by observational biases. 

The Subaru telescope carried out an unprecedented wide and deep imaging survey with Hyper Suprime-Cam
\citep[HSC;][]{miyazaki18}.
The Subaru Strategic Programme with HSC \citep[HSC-SSP;][]{aihara18} is composed of three layers: Wide
($\sim1,000\,\mathrm{deg^2}$, $i$-band depth of $m_i\sim26.0\,\mathrm{mag}$), Deep ($\sim25\,\mathrm{deg^2}$,
$m_i\sim26.5\,\mathrm{mag}$), and Ultradeep ($\sim3\,\mathrm{deg^2}$, $m_i\sim27.0\,\mathrm{mag}$).
Based on the first-year dataset of the Wide layer, \citet{toshikawa18} constructed a systematic sample ($N\sim200$)
of protocluster candidates at $z\sim4$ for the first time, which was used for various studies of clustering
analysis, correlation with QSOs \citep{uchiyama18}, infrared (IR) emission \citep{kubo19}, and brightest
protocluster members \citep{ito19,ito20}.
The same protocluster search technique was applied to the Deep/Ultradeep (DUD) layer to identify protoclusters
over a wider redshift range \citep{toshikawa24}.
In the DUD layer, deep $U$-band imaging is also available from the CFHT Large Area $U$-band Deep Survey
\citep[CLAUDS;][]{sawicki19}.
Following the identification of protocluster candidates at $z\sim3\mathrm{-}5$, this paper presents the results of
a pilot follow-up spectroscopy program in the DUD layer and investigates the structural and galaxy properties of a
spectroscopically confirmed protocluster. 

The paper is structured as follows.
Details of the target protocluster candidate and instrument configuration are outlined in Section~\ref{sec:obs}.
The results of the protocluster confirmation and characterisation of its properties are presented in
Section~\ref{sec:res}.
In Section~\ref{sec:dis}, we discuss a possible relation between cluster formation and galaxy evolution.
Finally, Section~\ref{sec:con} provides the conclusions of our pilot follow-up spectroscopy.
The following cosmological parameters are assumed: $\Omega_M=0.3$, $\Omega_\Lambda=0.7$, and
$H_0=70\,\mathrm{km\,s^{-1}\,Mpc^{-1}}$.
Magnitudes are given in the AB system.

\section{Observations} \label{sec:obs}
\subsection{Target} \label{sec:target}
Our target was drawn from a catalogue of protocluster candidates ($N\sim100$) from $z\sim3$ to $z\sim5$ in the
$25\,\mathrm{deg^2}$ area of the Deep/UltraDeep (DUD) layer in the HSC-SSP \citep{toshikawa24}.
Protocluster candidates were identified by the significant excess of surface number density of dropout galaxies,
which are commonly used and result in many discoveries of genuine protoclusters by follow-up spectroscopy
\citep[e.g.,][]{steidel98,douglas10,chanchaiworawit19,calvi21}. 
In \citet{toshikawa24}, surface overdensity is determined by the number of dropout galaxies within a fixed aperture
of $0.75\,\mathrm{pMpc}$ radius; then, $>4\sigma$ overdense regions are defined as protocluster candidates.
Analysing mock light cones based on theoretical models \citep{henriques15}, the purity of our sample of candidates
is found to be high ($\sim90\%$), at the expense of a low completeness ($\sim5\%\mathrm{-}10\%$).
It should be noted that, even in the regions including a genuine protocluster, many foreground or background
galaxies, which are not associated with a protocluster, are expected because the redshift window of dropout
selection ($\Delta z\sim1$) is about $\sim20$ times wider than the redshift size of a protocluster
($\Delta z\sim0.05$).
The number of interlopers will consequently be higher than that of protocluster members even in significantly
overdense regions.
Such high fraction of interlopers in an overdense region makes it difficult to conduct detailed studies on the
internal structure of an individual protocluster, or study subtle differences in physical properties of
protocluster members.
This prompted us to carry out follow-up spectroscopy to distinguish protocluster members from interlopers.

\begin{figure*}
\includegraphics[width=\hsize, bb=0 0 720 317]{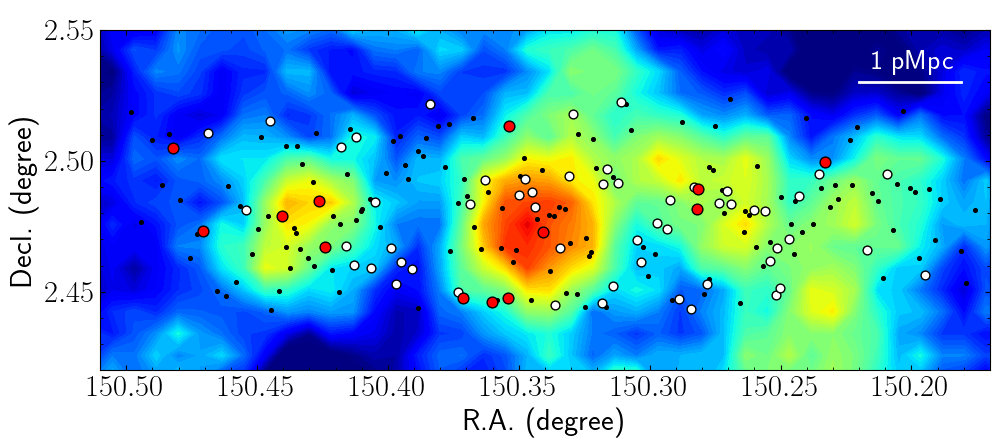}
\caption{Overdensity map of $g$-dropout galaxies around the ID4 protocluster candidate.
    Surface density is measured by counting $g$-dropout galaxies within a fixed aperture of $1.8\,\mathrm{arcmin}$
    radius, and higher density regions are indicated by redder colour.
    The overdensity at the peak is $4.5\sigma$.
    The red and white points are spectroscopically-confirmed protocluster members and foreground/background
    galaxies.
    The black dots are $g$-dropout galaxies which are observed by follow-up spectroscopy but do not exhibit (strong)
    Ly$\alpha$ emission.}
\label{fig:cntr}
\end{figure*}

Our catalogue of protocluster candidates is composed of $U$- to $r$-dropout galaxies, corresponding to star-forming
galaxies at $z\sim3\mathrm{-}5$.
We use Ly$\alpha$ emission to precisely determine their redshifts.
In addition, we focus on the COSMOS field because of the rich and deep multi-wavelength dataset that is available
over this area \citep[e.g.,][]{jin18,weaver22}.
Therefore, the ID4 protocluster candidate of $g$-dropout galaxies was selected as the target of our pilot follow-up
spectroscopy.
The target protocluster candidate has a $4.5\sigma$ overdensity significance \citep[see Table~2 in][]{toshikawa24}.
In the future, we will expand our campaign of follow-up spectroscopy to other protocluster candidates.

\subsection{Follow-up spectroscopy} \label{sec:spec}

Our spectroscopic observation was performed by Keck I\hspace{-1.2pt}I/DEIMOS \citep{faber03} on 17th January 2023
through the time exchange program with the Subaru telescope (proposal ID: S22B083).
We use the 900ZD grating which has a high efficiency over the target wavelength range of $5000\mathrm{-}6500$\,{\AA}
and sufficient spectral resolution to resolve the [\ion{O}{ii}] doublet at $z\sim0.5$, which is one of the major
contaminants in a spectroscopic observation.
DEIMOS combines a multi-object spectroscopy (MOS) mode with a wide field of view (FoV:
$16.7\times5.0\,\mathrm{arcmin^2}$).
The $g$-dropout galaxies in the ID4 protocluster candidate area are observed with two MOS masks, counting 115 and
95 slits per mask, respectively.
In addition to these science targets, one slit is allocated to a bright star ($\sim20\,\mathrm{mag}$) in each mask
to monitor sky conditions (e.g., atmospheric transparency and seeing size) between exposures.
The exposure time adopted for one shot was 20\,minutes, and the total integration time amounts to 200 and
180\,minutes for the first and second mask, respectively.
Figure~\ref{fig:cntr} shows the overdensity map of $g$-dropout galaxies and sky distribution of spectroscopic
targets.
The fraction of spectroscopically-observed galaxies among photometric $g$-dropout galaxies is $\sim30\%$ around
the overdensity peak and $\sim40\%$ on the outskirts.

We reduced the DEIMOS data using the PypeIt pipeline \citep{prochaska20}, which is a Python package for
semi-automated reduction of astronomical slit-based spectroscopic data.
On the combined two-dimensional spectra produced by PypeIt, possible emission lines are distinguished from fake
ones caused by sky residuals or ghosts by visual inspection.
Although PypeIt offers a function to extract one-dimensional spectra, we carried out the extractions manually in
order to optimally trace object positions, as the predicted position from the mask design could be shifted by up
to a few pixels.
For flux calibration, we have observed the spectroscopic standard stars G191B2B and HZ 44 at the beginning and end
of the observation, respectively.

\begin{figure*}
\includegraphics[width=\hsize, bb=0 0 720 790]{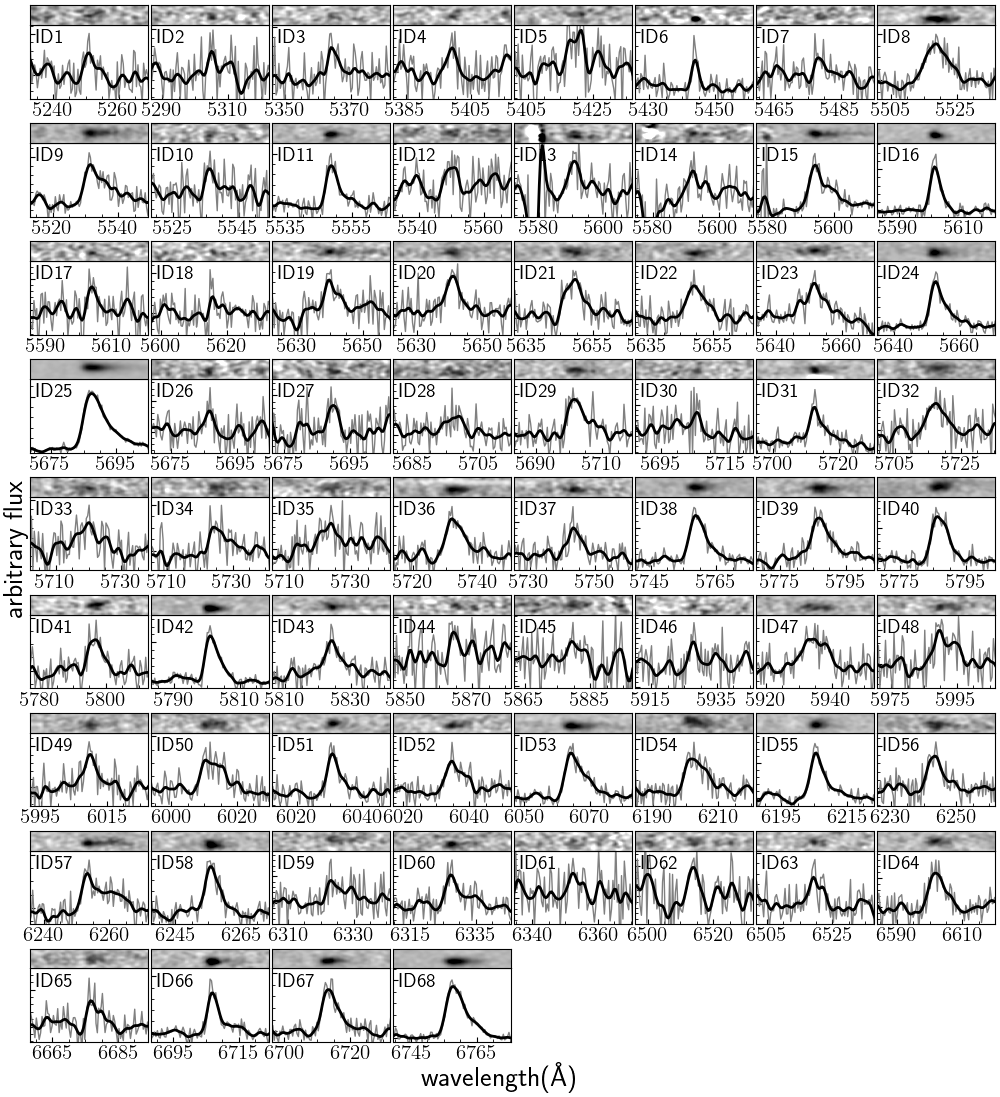}
\caption{One- and two-dimensional spectra of all $g$-dropout galaxies showing Ly$\alpha$ emission lines.
    The object IDs are indicated in the upper left corner (column~1 of Table~1).
    The gray and black one-dimensional spectra are raw and smoothed by a Gaussian with the spectral
    resolution as width, respectively.
    The dispersion is $0.44\,\mathrm{\text{\AA}\,pixel^{-1}}$ while the spectral resolution is
    $\Delta \lambda=2.8\,\text{\AA}$.}
\label{fig:mulsp}
\end{figure*}

We successfully identified emission lines from 68 galaxies among 210 spectroscopic targets.
Most emission lines are skewed redward, which is a clear spectroscopic signature of Ly$\alpha$ emission from
high-redshift galaxies (Figure~\ref{fig:mulsp}).
Such a skewed line profile can be quantified by weighted skewness, $S_w$ \citep{kashikawa06}, and most emission
lines are high, positive values.
No doublet like [\ion{O}{ii}] emission was identified, in spite of sufficient spectral resolution to detect it if
present (in which case the respective source would have been a low-redshift interloper).
Considering alternative low-redshift interlopers, the wide spectral coverage of DEIMOS would enable detection of
multiple emission lines if they were H$\beta$ or [\ion{O}{iii}].
Therefore, all single emission lines are regarded as Ly$\alpha$ emission.
Table~\ref{tab:spec} summarises the spectroscopic properties of the 68 confirmed galaxies.
As the UV continuum of most of our targets is too faint to be detected by our spectroscopy, UV absolute magnitude
($M_\mathrm{UV}$) and Ly$\alpha$ equivalent width in the rest frame ($EW_0$) are calculated by using $i$-band
photometry with the assumption of a flat UV continuum ($f_\lambda\propto\lambda^\beta$, where $\beta=-2$).
As for brightest $g$-dropout galaxies ($m_i\lesssim23.5$; ID15 and ID25), their UV continuum can be detected with
a signal-to-noise ratio of $S/N\sim2\mathrm{-}3$ by our spectroscopy.
Due to this low $S/N$, we derive UV magnitude based on the broad-band photometry, even for such bright galaxies.
Slit losses are assessed by using a bright star also observed within the same mask for calibration.

\section{Results} \label{sec:res}
\subsection{Protocluster confirmation} \label{sec:conf}
\begin{figure}
\includegraphics[width=\hsize, bb=0 0 364 292]{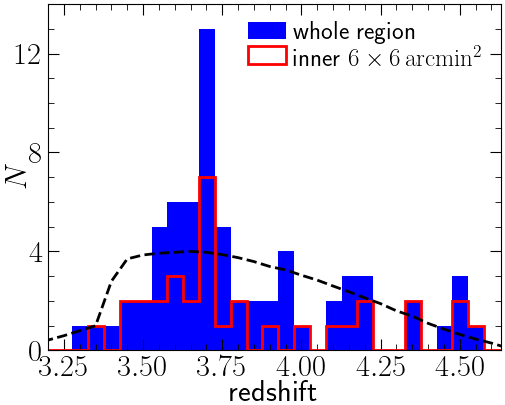}
\caption{Redshift distribution of spectroscopically-confirmed galaxies.
    The red line shows the subset of galaxies located near the peak in surface overdensity.
    The black dashed line indicates the selection function of $g$-dropout galaxies, which is scaled to the number
    of spectroscopically-confirmed $g$-dropout galaxies.
    There is a clear redshift peak at $z=3.699$.}
\label{fig:zhist}
\end{figure}

The redshift distribution of spectroscopically-confirmed galaxies exhibits a clear peak at $z=3.699$, including
thirteen galaxies within $\Delta z<0.05$ (Figure~\ref{fig:zhist}).
Based on the redshift window of $g$-dropout galaxies, the expected number of galaxies in the redshift bin of
$\Delta z=0.05$ is $N\sim4$.
According to Poisson statistics for such a small sample size \citep{gehrels86}, the probability of more than
thirteen galaxies clustering in such a narrow bin is only $<0.5\%$ by chance.
Therefore, the thirteen galaxies are expected to be physically associated and form a genuine structure.

\begin{table*}
\caption{Properties of spectroscopically-confirmed $g$-dropout galaxies.}
\label{tab:spec}
\begin{tabular}{cccccccccc}
\hline
ID & R.A. & Decl. & $m_i$ & Redshift & $M_\mathrm{UV}$ & $f_\mathrm{Ly\alpha}$ & $L_\mathrm{Ly\alpha}$ & $EW_0$ & $S_w$ \\
 & (J2000) & (J2000) & (mag) & & (mag) & ($10^{-18}\,\mathrm{erg\,s^{-1}\,cm^{-2}}$) & ($10^{42}\,\mathrm{erg\,s^{-1}}$) & (\AA) & (\AA) \\
\hline
1 & 10:01:39.87 & +02:28:02.7 & $26.99\pm0.13$ & $3.319\pm0.001$ & $-18.41\pm0.23$ & $3.78\pm1.34$ & $0.37\pm0.13$ & $18.3\pm7.8$ & $3.0\pm1.5$ \\
2 & 10:01:10.42 & +02:28:25.9 & $26.17\pm0.06$ & $3.363\pm0.001$ & $-19.36\pm0.11$ & $1.91\pm1.65$ & $0.19\pm0.17$ & $4.0\pm3.5$ & $3.6\pm3.5$ \\
3 & 10:01:00.45 & +02:26:55.5 & $24.87\pm0.02$ & $3.412\pm0.001$ & $-19.90\pm0.07$ & $2.41\pm1.47$ & $0.25\pm0.15$ & $3.2\pm1.9$ & $7.3\pm6.2$ \\
4 & 10:01:16.41 & +02:26:44.3 & $26.48\pm0.08$ & $3.441\pm0.001$ & $-19.16\pm0.13$ & $2.72\pm1.82$ & $0.29\pm0.20$ & $7.2\pm4.9$ & $0.0\pm20.9$ \\
5 & 10:01:07.96 & +02:29:23.7 & $26.05\pm0.05$ & $3.457\pm0.001$ & $-19.43\pm0.10$ & $3.79\pm1.49$ & $0.41\pm0.16$ & $7.9\pm3.2$ & $10.5\pm7.9$ \\
6 & 10:01:014.9 & +02:29:29.9 & $26.36\pm0.07$ & $3.478\pm0.001$ & $-19.18\pm0.13$ & $2.78\pm0.85$ & $0.31\pm0.09$ & $7.4\pm2.5$ & $2.7\pm1.5$ \\
7 & 10:01:08.22 & +02:26:37.1 & $25.23\pm0.02$ & $3.505\pm0.001$ & $-20.45\pm0.04$ & $1.90\pm1.11$ & $0.21\pm0.13$ & $1.6\pm0.9$ & $2.8\pm2.0$ \\
8 & 10:01:39.08 & +02:27:37.0 & $24.80\pm0.02$ & $3.540\pm0.001$ & $-20.91\pm0.03$ & $15.80\pm2.75$ & $1.81\pm0.32$ & $8.9\pm1.6$ & $3.4\pm5.5$ \\
9 & 10:01:49.05 & +02:28:53.0 & $24.47\pm0.01$ & $3.550\pm0.001$ & $-21.33\pm0.02$ & $26.36\pm2.83$ & $3.05\pm0.33$ & $10.2\pm1.1$ & $7.0\pm1.2$ \\
10 & 10:01:35.26 & +02:27:10.7 & $26.95\pm0.12$ & $3.554\pm0.001$ & $-18.60\pm0.22$ & $2.52\pm1.28$ & $0.29\pm0.15$ & $12.0\pm6.7$ & $2.0\pm2.1$ \\
11 & 10:01:19.07 & +02:31:04.8 & $27.03\pm0.13$ & $3.564\pm0.001$ & $-19.01\pm0.16$ & $10.72\pm1.51$ & $1.25\pm0.18$ & $35.2\pm7.5$ & $4.8\pm1.8$ \\
12 & 10:01:20.24 & +02:28:00.4 & $24.37\pm0.01$ & $3.566\pm0.001$ & $-21.01\pm0.03$ & $3.91\pm1.50$ & $0.46\pm0.18$ & $2.0\pm0.8$ & $3.6\pm2.3$ \\
13 & 10:01:40.26 & +02:30:19.1 & $25.95\pm0.05$ & $3.599\pm0.001$ & $-19.56\pm0.10$ & $5.30\pm1.87$ & $0.63\pm0.22$ & $10.7\pm3.9$ & $3.1\pm2.9$ \\
14 & 10:01:28.48 & +02:29:00.5 & $25.37\pm0.03$ & $3.600\pm0.001$ & $-20.35\pm0.05$ & $3.97\pm1.31$ & $0.48\pm0.16$ & $3.9\pm1.3$ & $1.8\pm1.8$ \\
15 & 10:01:32.12 & +02:31:18.8 & $23.29\pm0.00$ & $3.602\pm0.001$ & $-22.29\pm0.01$ & $26.05\pm2.59$ & $3.12\pm0.31$ & $4.3\pm0.4$ & $6.4\pm1.1$ \\
16 & 10:01:23.41 & +02:29:34.8 & $26.77\pm0.10$ & $3.608\pm0.001$ & $-19.10\pm0.15$ & $26.09\pm2.01$ & $3.14\pm0.24$ & $81.4\pm14.0$ & $4.7\pm1.5$ \\
17 & 10:00:59.22 & +02:28:13.0 & $25.98\pm0.05$ & $3.609\pm0.001$ & $-19.19\pm0.14$ & $1.47\pm0.91$ & $0.18\pm0.11$ & $4.2\pm2.7$ & $2.7\pm2.2$ \\
18 & 10:01:13.15 & +02:28:11.5 & $25.37\pm0.03$ & $3.619\pm0.001$ & $-20.30\pm0.05$ & $1.80\pm1.12$ & $0.22\pm0.14$ & $1.9\pm1.2$ & $1.3\pm1.0$ \\
19 & 10:01:52.54 & +02:30:38.7 & $25.41\pm0.03$ & $3.640\pm0.001$ & $-20.39\pm0.05$ & $5.36\pm1.37$ & $0.66\pm0.17$ & $5.2\pm1.3$ & $4.7\pm2.5$ \\
20 & 10:01:27.07 & +02:29:33.1 & $26.28\pm0.07$ & $3.640\pm0.001$ & $-19.73\pm0.09$ & $12.34\pm3.12$ & $1.52\pm0.38$ & $22.1\pm5.9$ & $3.4\pm2.8$ \\
21 & 10:01:05.63 & +02:29:02.1 & $25.15\pm0.02$ & $3.647\pm0.001$ & $-20.68\pm0.04$ & $14.65\pm2.10$ & $1.81\pm0.26$ & $10.9\pm1.6$ & $3.9\pm1.4$ \\
22 & 10:01:10.11 & +02:29:06.1 & $27.02\pm0.13$ & $3.647\pm0.001$ & $-18.80\pm0.21$ & $9.63\pm1.87$ & $1.19\pm0.23$ & $40.8\pm11.7$ & $3.2\pm2.2$ \\
23 & 10:01:00.06 & +02:27:04.8 & $25.79\pm0.04$ & $3.649\pm0.001$ & $-20.25\pm0.06$ & $11.03\pm1.67$ & $1.36\pm0.21$ & $12.3\pm2.0$ & $3.1\pm1.5$ \\
24 & 10:00:56.45 & +02:29:42.0 & $25.70\pm0.04$ & $3.650\pm0.001$ & $-20.14\pm0.07$ & $36.02\pm2.87$ & $4.45\pm0.36$ & $44.4\pm4.5$ & $7.5\pm0.8$ \\
25 & 10:01:45.73 & +02:28:44.8 & $23.53\pm0.01$ & $3.678\pm0.001$ & $-22.13\pm0.01$ & $123.72\pm3.43$ & $15.58\pm0.43$ & $24.9\pm0.7$ & $15.1\pm0.6$ \\
26 & 10:01:07.53 & +02:29:21.1 & $27.01\pm0.13$ & $3.678\pm0.001$ & $-18.64\pm0.24$ & $3.15\pm1.53$ & $0.40\pm0.19$ & $15.7\pm8.6$ & $0.6\pm1.7$ \\
27 & 10:00:55.94 & +02:29:59.1 & $26.02\pm0.05$ & $3.680\pm0.001$ & $-19.66\pm0.10$ & $2.29\pm1.33$ & $0.29\pm0.17$ & $4.5\pm2.6$ & $3.3\pm18.6$ \\
28 & 10:01:21.77 & +02:28:22.3 & $26.78\pm0.10$ & $3.686\pm0.001$ & $-18.80\pm0.21$ & $3.58\pm1.24$ & $0.45\pm0.16$ & $15.6\pm6.4$ & $0.7\pm1.9$ \\
29 & 10:01:41.81 & +02:28:01.3 & $25.40\pm0.03$ & $3.690\pm0.001$ & $-20.23\pm0.06$ & $7.20\pm1.55$ & $0.91\pm0.20$ & $8.4\pm1.9$ & $9.3\pm3.5$ \\
30 & 10:01:52.92 & +02:28:24.3 & $24.80\pm0.02$ & $3.693\pm0.001$ & $-20.82\pm0.04$ & $1.78\pm1.04$ & $0.23\pm0.13$ & $1.2\pm0.7$ & $-0.9\pm2.2$ \\
31 & 10:01:07.62 & +02:28:53.5 & $26.25\pm0.06$ & $3.699\pm0.001$ & $-19.55\pm0.11$ & $11.99\pm1.27$ & $1.53\pm0.16$ & $26.3\pm4.0$ & $3.8\pm0.7$ \\
32 & 10:01:26.43 & +02:26:46.0 & $24.62\pm0.01$ & $3.703\pm0.001$ & $-21.17\pm0.03$ & $8.56\pm2.28$ & $1.10\pm0.29$ & $4.2\pm1.1$ & $4.1\pm3.4$ \\
33 & 10:01:042.3 & +02:29:05.5 & $26.68\pm0.09$ & $3.705\pm0.001$ & $-19.27\pm0.15$ & $4.83\pm1.77$ & $0.62\pm0.23$ & $13.8\pm5.4$ & $1.8\pm1.7$ \\
34 & 10:01:29.14 & +02:26:51.5 & $25.23\pm0.03$ & $3.708\pm0.001$ & $-20.63\pm0.04$ & $9.03\pm2.62$ & $1.16\pm0.34$ & $7.4\pm2.2$ & $3.2\pm1.6$ \\
35 & 10:01:25.02 & +02:26:51.6 & $24.94\pm0.02$ & $3.708\pm0.001$ & $-20.74\pm0.04$ & $4.72\pm1.68$ & $0.61\pm0.22$ & $3.5\pm1.2$ & $1.3\pm3.5$ \\
36 & 10:01:24.86 & +02:30:48.0 & $26.26\pm0.06$ & $3.715\pm0.001$ & $-19.82\pm0.09$ & $20.89\pm2.42$ & $2.69\pm0.31$ & $36.0\pm5.2$ & $6.9\pm1.3$ \\
37 & 10:01:55.71 & +02:30:17.2 & $27.07\pm0.14$ & $3.724\pm0.001$ & $-18.94\pm0.20$ & $6.95\pm1.53$ & $0.90\pm0.20$ & $27.1\pm8.0$ & $4.0\pm1.1$ \\
38 & 10:01:020.7 & +02:26:41.3 & $26.13\pm0.06$ & $3.737\pm0.001$ & $-19.96\pm0.08$ & $31.06\pm2.59$ & $4.06\pm0.34$ & $47.8\pm5.5$ & $8.7\pm1.1$ \\
39 & 10:01:01.45 & +02:28:51.1 & $24.85\pm0.02$ & $3.759\pm0.001$ & $-20.87\pm0.04$ & $22.04\pm2.60$ & $2.92\pm0.34$ & $14.8\pm1.8$ & $7.9\pm1.2$ \\
40 & 10:01:00.34 & +02:27:59.8 & $24.32\pm0.01$ & $3.759\pm0.001$ & $-21.41\pm0.02$ & $33.57\pm3.12$ & $4.45\pm0.41$ & $13.8\pm1.3$ & $7.8\pm1.3$ \\
41 & 10:01:37.51 & +02:27:32.9 & $27.13\pm0.14$ & $3.767\pm0.001$ & $-18.86\pm0.22$ & $7.11\pm1.66$ & $0.95\pm0.22$ & $30.7\pm9.9$ & $2.3\pm1.2$ \\
42 & 10:01:37.21 & +02:29:04.1 & $26.54\pm0.08$ & $3.772\pm0.001$ & $-19.40\pm0.14$ & $18.22\pm1.45$ & $2.44\pm0.19$ & $48.2\pm7.6$ & $6.9\pm1.3$ \\
43 & 10:01:24.02 & +02:29:12.8 & $26.67\pm0.09$ & $3.791\pm0.001$ & $-19.44\pm0.14$ & $12.37\pm2.41$ & $1.67\pm0.33$ & $31.9\pm7.6$ & $4.1\pm1.3$ \\
44 & 10:01:29.59 & +02:27:00.4 & $26.74\pm0.10$ & $3.824\pm0.001$ & $-19.31\pm0.16$ & $1.52\pm1.17$ & $0.21\pm0.16$ & $4.5\pm3.5$ & $5.0\pm8.9$ \\
45 & 10:01:34.77 & +02:27:41.0 & $24.57\pm0.01$ & $3.837\pm0.001$ & $-21.33\pm0.03$ & $4.83\pm1.59$ & $0.67\pm0.22$ & $2.2\pm0.7$ & $1.1\pm2.1$ \\
46 & 10:00:046.8 & +02:27:23.7 & $26.24\pm0.06$ & $3.876\pm0.001$ & $-19.08\pm0.20$ & $2.60\pm0.94$ & $0.37\pm0.13$ & $9.8\pm4.1$ & $3.8\pm1.7$ \\
47 & 10:01:11.29 & +02:28:34.2 & $25.73\pm0.04$ & $3.882\pm0.001$ & $-19.95\pm0.10$ & $14.01\pm2.49$ & $2.01\pm0.36$ & $23.8\pm4.8$ & $8.7\pm3.6$ \\
48 & 10:00:50.25 & +02:29:41.5 & $25.79\pm0.04$ & $3.926\pm0.001$ & $-20.15\pm0.08$ & $7.14\pm2.01$ & $1.05\pm0.30$ & $10.3\pm3.0$ & $6.5\pm3.8$ \\
49 & 10:00:52.04 & +02:27:57.1 & $26.89\pm0.11$ & $3.943\pm0.001$ & $-19.29\pm0.18$ & $11.27\pm1.58$ & $1.67\pm0.24$ & $36.4\pm8.3$ & $4.9\pm1.0$ \\
50 & 10:01:02.44 & +02:28:52.8 & $26.17\pm0.06$ & $3.945\pm0.001$ & $-19.57\pm0.14$ & $9.81\pm2.65$ & $1.46\pm0.39$ & $24.5\pm7.5$ & $7.4\pm4.5$ \\
51 & 10:00:58.28 & +02:29:12.3 & $27.20\pm0.15$ & $3.961\pm0.001$ & $-18.82\pm0.27$ & $8.55\pm1.47$ & $1.28\pm0.22$ & $43.3\pm14.4$ & $3.3\pm2.1$ \\
52 & 10:01:35.74 & +02:28:01.0 & $27.18\pm0.15$ & $3.964\pm0.001$ & $-18.73\pm0.29$ & $6.69\pm1.59$ & $1.01\pm0.24$ & $36.9\pm14.4$ & $2.7\pm1.8$ \\
53 & 10:01:12.79 & +02:27:40.4 & $24.78\pm0.02$ & $3.989\pm0.001$ & $-21.14\pm0.04$ & $32.40\pm2.70$ & $4.95\pm0.41$ & $19.7\pm1.8$ & $8.9\pm1.8$ \\
54 & 10:01:06.75 & +02:27:11.1 & $26.63\pm0.09$ & $4.102\pm0.001$ & $-19.61\pm0.16$ & $18.10\pm2.89$ & $2.95\pm0.47$ & $48.0\pm11.0$ & $9.5\pm3.9$ \\
55 & 10:01:04.92 & +02:29:18.6 & $26.37\pm0.07$ & $4.104\pm0.001$ & $-19.68\pm0.15$ & $20.20\pm1.76$ & $3.30\pm0.29$ & $50.3\pm8.8$ & $6.8\pm0.8$ \\
56 & 10:01:46.84 & +02:30:55.3 & $26.26\pm0.06$ & $4.136\pm0.001$ & $-19.94\pm0.13$ & $18.28\pm3.52$ & $3.04\pm0.59$ & $36.7\pm8.5$ & $5.7\pm2.5$ \\
57 & 10:01:14.63 & +02:31:21.1 & $24.54\pm0.01$ & $4.144\pm0.001$ & $-21.43\pm0.03$ & $14.81\pm2.02$ & $2.48\pm0.34$ & $7.5\pm1.1$ & $5.0\pm1.3$ \\
58 & 10:01:33.76 & +02:27:32.1 & $26.60\pm0.09$ & $4.146\pm0.001$ & $-19.56\pm0.18$ & $15.44\pm1.88$ & $2.58\pm0.32$ & $43.8\pm9.4$ & $7.6\pm1.7$ \\
59 & 10:01:00.94 & +02:27:41.8 & $25.84\pm0.04$ & $4.201\pm0.001$ & $-20.48\pm0.08$ & $2.91\pm1.05$ & $0.50\pm0.18$ & $3.7\pm1.4$ & $2.4\pm1.5$ \\
60 & 10:01:15.88 & +02:29:49.4 & $27.01\pm0.13$ & $4.205\pm0.001$ & $-19.59\pm0.18$ & $5.67\pm1.30$ & $0.98\pm0.22$ & $16.3\pm4.8$ & $3.7\pm3.0$ \\
61 & 10:01:15.39 & +02:27:08.6 & $27.04\pm0.13$ & $4.225\pm0.001$ & $-19.33\pm0.23$ & $2.13\pm1.20$ & $0.37\pm0.21$ & $7.8\pm4.8$ & $1.3\pm1.6$ \\
62 & 10:01:22.81 & +02:29:17.5 & $25.11\pm0.02$ & $4.358\pm0.001$ & $-21.45\pm0.04$ & $3.08\pm1.07$ & $0.58\pm0.20$ & $1.7\pm0.6$ & $4.1\pm5.9$ \\
63 & 10:01:09.29 & +02:26:50.6 & $27.56\pm0.21$ & $4.363\pm0.001$ & $-19.51\pm0.24$ & $2.72\pm1.00$ & $0.51\pm0.19$ & $9.2\pm4.0$ & $0.7\pm1.0$ \\
\hline
\end{tabular}
\end{table*}

\begin{table*}
\contcaption{}
\begin{tabular}{cccccccccc}
\hline
ID & R.A. & Decl. & $m_i$ & Redshift & $M_\mathrm{UV}$ & $f_\mathrm{Ly\alpha}$ & $L_\mathrm{Ly\alpha}$ & $EW_0$ & $S_w$ \\
 & (J2000) & (J2000) & (mag) & & (mag) & ($10^{-18}\,\mathrm{erg\,s^{-1}\,cm^{-2}}$) & ($10^{42}\,\mathrm{erg\,s^{-1}}$) & (\AA) & (\AA) \\
\hline
64 & 10:01:04.53 & +02:29:00.2 & $26.04\pm0.05$ & $4.431\pm0.001$ & $-20.35\pm0.13$ & $9.03\pm1.87$ & $1.77\pm0.37$ & $14.5\pm3.5$ & $4.9\pm3.1$ \\
65 & 10:01:22.47 & +02:28:56.1 & $26.95\pm0.12$ & $4.492\pm0.001$ & $-19.43\pm0.30$ & $5.25\pm1.46$ & $1.06\pm0.30$ & $20.4\pm8.7$ & $3.2\pm1.4$ \\
66 & 10:01:16.32 & +02:29:28.7 & $27.28\pm0.16$ & $4.517\pm0.001$ & $-19.43\pm0.31$ & $9.29\pm1.14$ & $1.90\pm0.23$ & $36.3\pm13.0$ & $4.6\pm0.7$ \\
67 & 10:01:38.96 & +02:30:32.8 & $26.52\pm0.08$ & $4.523\pm0.001$ & $-19.65\pm0.27$ & $13.85\pm1.78$ & $2.85\pm0.37$ & $44.4\pm13.7$ & $5.5\pm3.4$ \\
68 & 10:01:19.38 & +02:29:39.5 & $25.91\pm0.05$ & $4.559\pm0.001$ & $-20.49\pm0.14$ & $41.50\pm1.62$ & $8.69\pm0.34$ & $62.8\pm9.0$ & $7.5\pm0.7$ \\
\hline
\end{tabular}
\end{table*}

\begin{figure}
\includegraphics[width=\hsize, bb=0 0 360 432]{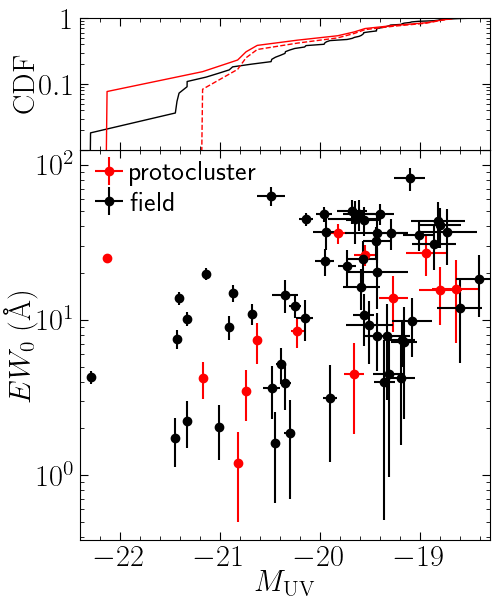}
\caption{Physical properties of spectroscopically-confirmed $g$-dropout galaxies ($M_\mathrm{UV}$ and $EW_0$).
    The upper panel shows the cumulative distribution function (CDF) for protocluster and field galaxies.
    The dashed red line is derived by removing the brightest protocluster member, or AGN.}
\label{fig:MuvEW}
\end{figure}

The structure at $z=3.699$ found by our spectroscopy is composed of thirteen galaxies at least, which is comparable
to known protoclusters, and the galaxies are distributed over $\sim12\,\mathrm{arcmin}$ corresponding to
$\sim5.2\,\mathrm{pMpc}$.
The size of protoclusters is tightly correlated with the descendant halo mass at $z=0$ \citep{chiang13,muldrew15}.
If our newly found structure is the progenitor of a rich cluster ($M_h>10^{15}\,\mathrm{M_{\sun}}$), the spatial
distribution of its member galaxies at $z\sim3\mathrm{-}4$ can indeed be expected to be on the order of
$\sim4\mathrm{-}6\,\mathrm{pMpc}$.
On the other hand, considering the decreasing abundance of galaxy clusters with increasing mass, one may expect
progenitors of typical galaxy clusters ($M_h\sim10^{14}\,\mathrm{M_{\sun}}$) to be more frequently discovered than
that of rich clusters.
The size of the progenitors of such typical galaxy clusters is $\sim2\mathrm{-}3\,\mathrm{pMpc}$.
In contrast, previous studies confirmed several rich and large protoclusters: \citet{lemaux18} reported that nine
spectroscopic protocluster members at $z=4.57$ are spread across an aperture of $\sim6\,\mathrm{pMpc}$ diameter,
a giant protocluster at $z=5.7$ having $\sim5\,\mathrm{pMpc}$ size was found by \citet{linhua18}, and a complex of
protoclusters extending over $\sim20\,\mathrm{pMpc}$ at $z=2.45$ was also discovered by \citet{cucciati18}.
Despite their rarity, it may be more feasible to confirm such rich protoclusters in actual observations.
It should be noted that the shape of protoclusters at high redshifts is far from spherical, with filamentary
elongations tracing the cosmic web \citep{lovell18}. 
Thus, the length of major axes of rich protoclusters can be $\sim8\,\mathrm{pMpc}$. 
As shown in Figure~\ref{fig:cntr}, the overdense region of the ID4 protocluster candidate expands into the west
and east directions.
Since the FoVs of both DEIMOS masks were deliberately aligned with the elongation of the overdense region, wider
coverage of the area surrounding the ID4 protocluster candidates (e.g., northern and southern parts) would be
required in order to unbiasedly reveal the spatial distribution of member galaxies.
Although it is difficult to draw a clear conclusion on its descendant structure at $z=0$, the thirteen galaxies
clustering at $z=3.699$ can be regarded as a protocluster, possibly the progenitor of a rich cluster.

\subsection{Galaxy properties} \label{sec:prop}

We next compare the galaxy properties between the thirteen protocluster members and field galaxies.
As the field galaxies are identified as foreground/background galaxies from the same spectroscopic observation,
there is less observational bias between the two samples.
As shown in Figure~\ref{fig:MuvEW}, the overall trend of $M_\mathrm{UV}$ and $EW_0$ of protocluster members is not
significantly different from that of field counterparts.
That is, dropout galaxies with brighter UV magnitudes tend to have smaller $EW_0$ on average, regardless of their
protocluster or field status.
Such a trend is generally interpreted as more massive (and therefore brighter) SFGs being dustier, making it harder
for the Ly$\alpha$ emission to escape.
The fact that Ly$\alpha$ suffers more from dust attenuation than the UV continuum stems from its resonant
scattering nature \citep[see, e.g.,][for a discussion on Ly$\alpha$ radiative transfer]{verhamme06,blaizot23}.

\begin{figure}
\includegraphics[width=\hsize, bb=0 0 432 288]{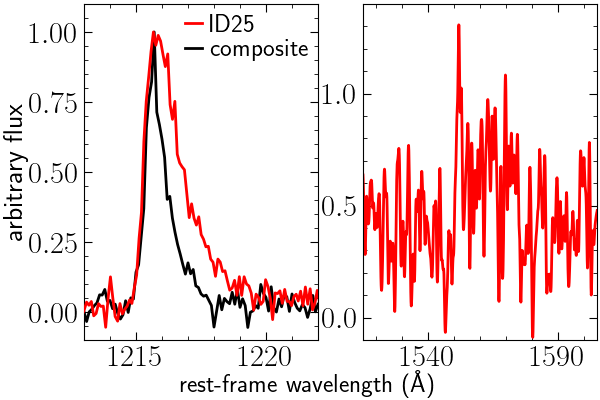}
\caption{Spectrum of the brightest protocluster member zoomed in on Ly$\alpha$ and possible \ion{C}{iv} emission
    in the left and right panels, respectively.
    For comparison, the composite Ly$\alpha$ emission line of the other $g$-dropout galaxies confirmed by the
    follow-up spectroscopy of this study are also plotted, enabling us to notice the broad Ly$\alpha$ emission of
    the brightest protocluster member.}
\label{fig:sp1dAGN}
\end{figure}

In detail, some signs are present that set the protocluster members apart from field galaxies, especially at the
bright end of the UV luminosity distribution.
The fraction of particularly bright galaxies ($M_\mathrm{UV}<-22$) is only $\sim0.02$ in the field; nevertheless,
one of the thirteen protocluster members has such brightness.
Given the bright-galaxy fraction of 0.02 ($1/55$) derived from field galaxies, the expected number of bright
galaxies among the thirteen protocluster members is $0.24\pm0.23$ according to a binomial distribution.
This tendency of a relatively more populated UV-luminous end was already suggested by the photometric study of
\citet{toshikawa24}, and the follow-up spectroscopy shows a consistent result.
This implies that there is no significant difference between the fraction of Ly$\alpha$ emitters among dropout
galaxies in the protocluster and the field.
Moreover, we note that in this case the brightest protocluster member deviates from the general relation between
$M_\mathrm{UV}$ and $EW_0$, and its Ly$\alpha$ emission is found not only to be intense but also broad as shown in
the left panel of Figure~\ref{fig:sp1dAGN}.
The full width at half maximum of its Ly$\alpha$ emission is $\mathrm{FWHM}=1.7\,\text{\AA}$ in the rest frame or
$\mathrm{FWHM}=420\,\mathrm{km\,s^{-1}}$.
The brightest protocluster member has another emission line at $\lambda\sim7300\,\text{\AA}$, corresponding to
$\sim1560\,\text{\AA}$ in the rest frame.
Its integrated $S/N$ over $\lambda_\mathrm{rest}=1550\mathrm{-}1575\,\text{\AA}$ is $S/N\sim7$.
Although this wavelength is slightly different from that of \ion{C}{iv} emission ($\lambda=1549.05\,\text{\AA}$),
the observed wavelength of Ly$\alpha$ or \ion{C}{iv} emissions could be shifted from the galaxy's systemic redshift
due to galactic kinematics \citep[see, e.g.,][]{rankine20}.
Furthermore, the kinematic state of the broad line region of AGNs can cause large line profile differences for
high-ionisation lines like \ion{C}{iv} from object to object, and some AGNs exhibit absorption features superposed
on a broad \ion{C}{iv} emission line.
Strong OH sky lines around $\lambda\sim7300\,\text{\AA}$ can also hamper accurate line profile measurement.
Thus, the emission line at $\lambda\sim7300\,\text{\AA}$ can be regarded as \ion{C}{iv} emission.
Both intense/broad Ly$\alpha$ emission and the higher order ionisation emission of \ion{C}{iv} can be attributed to
the presence of an active galactic nucleus (AGN).
As for field galaxies, no clear AGN is found.
The bright-end excess is ascribed to the AGN, though there is still a marginal excess around
$M_\mathrm{UV}\sim-20.5\,\mathrm{mag}$ (the upper panel of Figure~\ref{fig:MuvEW}).

In contrast to the brightest protocluster member, the other bright protocluster members tend to have smaller $EW_0$
than field galaxies.
The median $\log(EW_0/\text{\AA})$ of protocluster members at $-22<M_\mathrm{UV}<-20$ is
$\log(EW_0/\text{\AA})=0.63\pm0.30$, compared to $\log(EW_0/\text{\AA})=0.98\pm0.45$ for field counterparts.
As Ly$\alpha$ emission is sensitive to dust attenuation, its strength relative to the UV continuum can be taken as
a proxy for ISM maturity.
The sense of the offset would then convey that brighter protocluster members may be more mature compared to field
counterparts, albeit not at a statistically significant level. 
We have also checked the distribution of $EW_0$, and there is no significant difference between protocluster
members and field galaxies ($p$-value of Kolmogorov-Smirnov test is $p=0.14$).

\subsection{Protocluster structure} \label{sec:struc}

The three-dimensional (3D) distribution of protocluster members can be regarded as one of the key parameters
marking the developmental phase of galaxy clusters and predictive of environmental effects on galaxy evolution.
The evolution of size or internal structure of protoclusters are investigated by several theoretical works
\citep[e.g.,][]{chiang13,muldrew15,lovell18}.
As protoclusters are maturing, their shape, or the spatial distribution of member galaxies, tends to become more
spherical, after an earlier developmental stage characterized by elongated or irregular shapes.
In this study, a robust assessment of the shape of the protocluster is hampered by the limited FoV of our follow-up
spectroscopy and the selection bias of protocluster members.

Here, we just calculate distance from the protocluster centre, $D_\mathrm{cen}$, and separation from the nearest
neighbour, $D_\mathrm{nei}$, based on the thirteen spectroscopically-confirmed protocluster members.
The protocluster centre is simply defined by the average position of protocluster members.
This estimate of a centre assumes that Ly$\alpha$-emitting galaxies are unbiasedly distributed in the protocluster.
Although it is quite difficult to directly check the validity of this assumption, Ly$\alpha$-emitting galaxies tend
to be low-mass, young galaxies, implying that they are a major galaxy population in terms of number in a
protocluster.
We use redshifts as an indicator of line-of-sight distance as peculiar motions have only a minor effect on the
observed redshifts.
According to  \citet{henriques15}, the standard deviation of the difference between redshifts with and without
peculiar motion is only $3.1\times10^{-3}$ for $g$-dropout galaxies.
Combined with the spectral resolution of our follow-up spectroscopy ($\mathrm{FWHM}=2.8\,\text{\AA}$ or
$\delta z=9.8\times10^{-4}$), the error in cosmological redshift is estimated to be $\delta z=3.3\times10^{-3}$.
This redshift uncertainty corresponds to $0.53\,\mathrm{pMpc}$.
The error on the sky position (R.A. and Decl.) is negligible since the imaging dataset of HSC-SSP has a sufficiently
high astrometric accuracy (standard deviation of $\sim0.01\mathrm{-}0.02\,\mathrm{arcsec}$).
We quantify the errors on the structural parameters $D_\mathrm{cen}$ and $D_\mathrm{nei}$ using Monte Carlo
simulations.
Specifically, the redshifts (line-of-sight distance) of protocluster members are perturbed according to a Gaussian
distribution with $\sigma=3.3\times10^{-3}$ ($0.53\,\mathrm{pMpc}$), and for each Monte Carlo realisation the
structural parameters are calculated based on the perturbed redshifts.
This process is repeated 10,000 times.
The lower and upper errors are determined by the 16th and 84th percentiles, respectively.
If redshift difference from the centre of the protocluster (the nearest neighbor) is almost zero, perturbing the
redshift causes only larger separations.
As for $D_\mathrm{nei}$, when a protocluster member is located near the middle of two other members, it is rare
that $D_\mathrm{nei}$ becomes larger by perturbing redshifts.  This explains the sometimes asymmetric error bars.

\begin{figure}
\includegraphics[width=\hsize, bb=0 0 385 346]{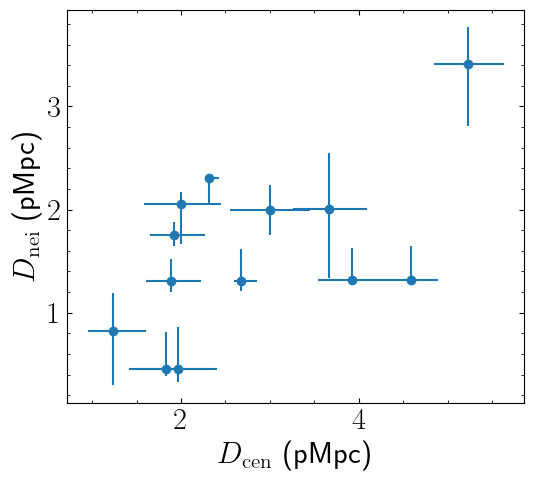}
\caption{Relation between $D_\mathrm{cen}$ and $D_\mathrm{nei}$ (distance to protocluster centre and to closest
    other protocluster member) of thirteen protocluster members.}
\label{fig:DcenDnei}
\end{figure}

We show the relation between $D_\mathrm{cen}$ and $D_\mathrm{nei}$ in Figure~\ref{fig:DcenDnei}.
The estimate of the centre of the protocluster is less sensitive to the number of confirmed members, while the
absolute separation from the nearest neighbor directly depends on the number density of confirmed members.
If the completeness of our follow-up spectroscopy is uniform over the FoV, the separation from the nearest
neighbour is still meaningful from a relative viewpoint.
Protocluster members near the centre tend to have smaller $D_\mathrm{nei}$ ($\rho=0.59$ and $p=0.03$ based on the
Spearman's rank-order correlation test).
Although observational biases (e.g., FoV of MOS masks and completeness) could have a significant effect on the
measurement of 3D structure, the positive correlation between $D_\mathrm{cen}$ and $D_\mathrm{nei}$ indicates a
centrally concentrated distribution of protocluster members.
Strong substructure with galaxy groups falling onto one another would leave a lesser correlation
\citep[see, e.g.,][for an example]{toshikawa20}.

\section{Discussion} \label{sec:dis}

\subsection{Environmental dependence of galaxy properties} \label{sec:env}

\begin{figure}
\includegraphics[width=\hsize, bb=0 0 602 555]{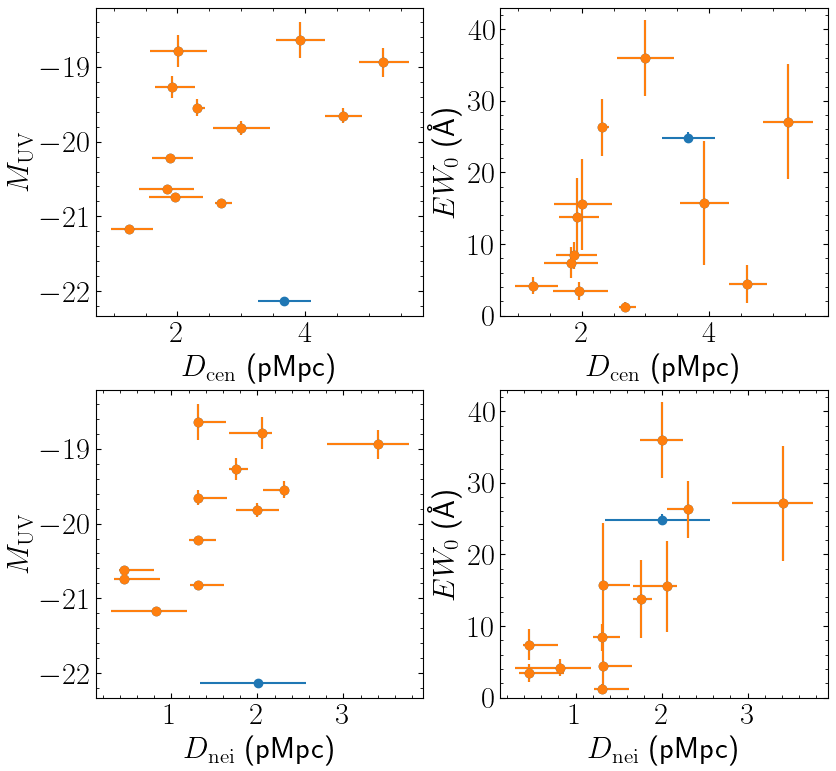}
\caption{Relations between structural parameters (top panels for $D_\mathrm{cen}$, and bottom ones for
    $D_\mathrm{nei}$) and galaxy properties (left panels for $M_\mathrm{UV}$, and right ones for Ly$\alpha$ $EW_0$).
    The AGN is represented by blue points.}
\label{fig:DsGal}
\end{figure}

Informed by the above results on galaxy properties and 3D structure of the protocluster, we now discuss the
relation between galaxy evolution and cluster formation. 
Figure~\ref{fig:DsGal} shows the relations between structural parameters ($D_\mathrm{cen}$ and $D_\mathrm{nei}$)
and galaxy properties ($M_\mathrm{UV}$ and $EW_0$).
There are tight correlations between $D_\mathrm{nei}$ and galaxy properties while $D_\mathrm{cen}$ is weakly
correlated.
The correlation coefficient and $p$-value for $D_\mathrm{nei}$ and $M_\mathrm{UV}$ ($EW_0$) are $\rho=0.74$ (0.81)
and $p<0.01$, those for $D_\mathrm{cen}$ are $\rho=0.54$ (0.43) and $p=0.07$ (0.16).
Here, we excluded the AGN (blue point) from the analysis because the physical mechanism of its strong UV/Ly$\alpha$
emission is entirely different from that in normal star-forming galaxies.
Considering the fact that $D_\mathrm{cen}$ and $D_\mathrm{nei}$ are correlated among themselves, it cannot be
denied that $D_\mathrm{cen}$ and galaxy properties are indirectly correlated through $D_\mathrm{nei}$.

These results can be interpreted as the protocluster still being in an early phase of cluster formation where
individual members are dependent on their local environments rather than the global structure of their host
protocluster.
This supports a hierarchical, or bottom-up, picture for structure formation.
Protoclusters still experience drag from cosmic expansion and their physical size is increasing; thus, at high
redshifts, protocluster members can have a physical connection only with their close neighbours.
Smaller $D_\mathrm{nei}$, or higher local density, could result in a larger amount of gas accretion and/or more
frequent galaxy mergers, and galaxy evolution or star-forming activity is enhanced as expected by brighter UV
magnitude.
Similarly, a reversal of the SFR-density relation at high redshifts ($2\lesssim z \lesssim5$) is reported by
several works \citep[e.g.,][]{lemaux22,taamoli23,shi24}.
Although this study cannot reveal the underlying physical mechanism, it illustrates the apparent relation between
environments and galaxy properties is dependent on the choice of the environmental indicator used since
environmental effects on galaxy evolution, in general, involve a wide range of physics not only for enhancing but
also quenching star-forming activity \citep[e.g.,][]{somerville15,alberts22}.

\begin{figure*}
\includegraphics[width=\hsize, bb=0 0 720 317]{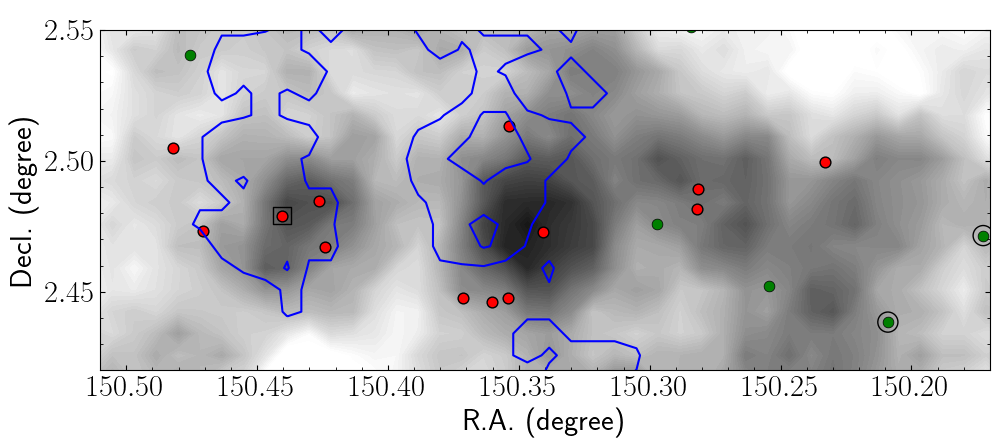}
\caption{Overdensity map of $g$-dropout galaxies (grayscales) and the sky distribution of confirmed  protocluster
    members (red circles) as shown in Figure~\ref{fig:cntr}.
    In addition, the blue lines are the ($2\sigma$ and $4\sigma$) overdensity contours of massive galaxies selected
    from COSMOS2020, and the green points are dusty star-burst galaxies from the COSMOS Super-deblended catalogue.
    The AGN and super-deblended sources with spectroscopic redshifts are shown by
    black square and circles, respectively.}
\label{fig:C20SD}
\end{figure*}

The environment indicator of $D_\mathrm{nei}$ traces a relatively smaller scale than $D_\mathrm{cen}$; however, it
cannot distinguish an isolated galaxy from a post-merger galaxy despite the fact that the underlying physical
situations are completely different.
The presence of an AGN in the protocluster may provide a clue on this.
If AGN activity is triggered by galaxy mergers \citep[e.g.,][]{hopkins08,byrne24}, one may expect on
statistical grounds that the AGN would most likely occur in the locally highest-density region in the protocluster.
As illustrated in Figure~\ref{fig:DsGal}, the AGN is located neither near the centre of the protocluster nor in the
neighbourhood of smaller-$D_\mathrm{nei}$ galaxies.
Instead, the location of the AGN is on the outskirt of the protocluster, which is a less probable outcome in a
scenario where nuclear activity is triggered by mergers.
Of course, AGNs being rare events and this study investigating only a single protocluster makes it hard to
generalize any trends discerned.
In another protocluster, discovered at $z=3.721$ following the same method as employed here, \citet{toshikawa20}
found the single AGN presents to be located in the central part of the protocluster for example. 
\citet{noirot18} confirmed 16 protoclusters by targeting RGs as signposts, but RGs are not always located at the
centre of the structures.
Extending such investigation to other known protoclusters would lead to a more heterogeneous sample as they were
discovered by various methods, thus hampering a fair comparison.
The HSC-SSP provides a systematic sample of protocluster candidates across cosmic time
\citep{toshikawa18,higuchi19,toshikawa24}.
Combined with the campaign of follow-up spectroscopy, we will be able to draw a conclusion on the relation between
protoclusters and rare objects like AGNs in the future.

\subsection{Other galaxy populations} \label{sec:pub}

We have confirmed the correlation between $D_\mathrm{nei}$ and galaxy properties, indicating that locally
high-density environments enhance star-forming activity.
This can most plausibly be read as galaxy evolution being accelerated in dense environments, although an
enhancement of starburst triggering may contribute as well.
Our method relies on $g$-dropout galaxies, and thus focuses on the dust-poor star-forming galaxy population, even
more so given that Ly$\alpha$ emission is needed to pinpoint the $g$-dropout galaxies in 3D space.
Such selection may miss the most vigorous star-forming systems, which tend to be enshrouded by dust, or other
objects with high mass-to-light ratios.
Actually, some protoclusters even at $z\gtrsim4$ are found to host massive quiescent galaxies 
\citep{tanaka23,kakimoto24}.
In order to find relatively mature galaxies or starburst galaxies, if any, we make use of the public COSMOS2020
\citep{weaver22} and Super-deblended \citep{jin18} catalogues.
Even a deep and well-sampled multiwavelength dataset such as available in the COSMOS field is unable to
photometrically constrain redshifts with an accuracy equivalent to the redshift size of the protocluster
($\Delta z<0.05$).
Therefore, we simply select galaxies whose best photometric redshifts are within the interval
$3.675<z_\mathrm{phot}<3.725$, corresponding to the redshift range of the protocluster, irrespective of the size
of their photo-z error.
As the typical error is $\sigma_{z_\mathrm{phot}}\sim0.7$ for $z\sim3.7$ galaxies in the COSMOS2020 catalogue, it
is meaningless to judge whether individual galaxies are protocluster members or not.
However, despite the unavoidable dilution by interlopers, it remains useful to evaluate whether an enhancement of
mature galaxies is present in the vicinity of the protocluster.

Relatively mature galaxies can be selected from the COSMOS2020 catalogue by $\log(M_\mathrm{star})>9.0$.
This threshold accounts for about the upper half of more massive galaxies among all galaxies at
$z_\mathrm{phot}\sim3.7$.
As there is a correlation between stellar mass and age, it is highly expected that more massive galaxies are
equivalent to older, more mature galaxies.
According to the CLASSIC catalogue with LePhare photometric redshifts, $N=1382$ galaxies are selected over the
whole survey area of COSMOS2020, resulting in 2.4 such galaxies per 1\farcm8-radius aperture on average.
Having calculated the surface overdensity of these massive galaxies by the same method as we applied for
$g$-dropout galaxies in \citet{toshikawa24}, we find massive galaxies at $z=3.7$ to also show a high overdensity
near the peak overdensity of $g$-dropout galaxies (Figure~\ref{fig:C20SD}).
This suggests that massive galaxies densely populate the core of the protocluster.
It is not common for such mature galaxies to have strong Ly$\alpha$ emissions; thus, the lack of confirmed 
protocluster members within $D_\mathrm{cen}<1\,\mathrm{pMpc}$ (Figure~\ref{fig:DsGal}) could be attributed to the
observational bias of our follow-up spectroscopy, which entirely relies on the presence of (strong) Ly$\alpha$
emission.
Although the existence of the protocluster was confirmed by our follow-up spectroscopy targeting Ly$\alpha$
(Section~\ref{sec:conf}), the protocluster is expected to be composed of multiple galaxy populations based on the
sky distribution of mature galaxies.
Another interesting feature of the overdensity map of mature galaxies is that the eastern part of the protocluster
shows moderate overdensity while there is no overdensity in the western part.
The AGN also happens to be located in the eastern part.
Perhaps, the eastern part is a subgroup falling into the main component of the protocluster.

The number of galaxies contained in the Super-deblended catalogue, which contains far-infrared photometry, is about
a tenth of that in the COSMOS2020 catalogue.
This is too scarce to measure overdensity.
Thus, galaxies at $z=3.700\pm0.025$ selected from the Super-deblended catalogue are just overplotted on the
overdensity map of $g$-dropout galaxies in Figure~\ref{fig:C20SD}.
There are five galaxies around the protocluster, and they seem to be biased toward the western part.
Their SFRs are inferred to be $\sim250\mathrm{-}1200\,\mathrm{M_{\sun}\,yr^{-1}}$ from the combination of optical
to far-infrared photometry and SED fitting technique, which is $\times10\mathrm{-}100$ higher than the typical SFR
of dropout galaxies.
It should be noted that two of the five infrared-detected galaxies were already confirmed to be at $z=3.709$ and
3.710 by spectroscopy \citep{lilly09,tasca17}.
Their redshifts are perfectly matched to the redshift range of the protocluster.
The total SFR of these two Super-deblended sources with spectroscopic redshifts is
$\sim1000\,\mathrm{M_{\sun}\,yr^{-1}}$.
Similar to the AGN, the dusty star-burst galaxies are located in the outskirts of the protocluster, provided that
$g$-dropout (i.e., dust-poor star-forming) galaxies correctly trace the protocluster structure.
It should be noted that there are some clear examples of protocluster cores composed of dusty star-burst galaxies
at $z\sim4$ \citep{miller18,oteo18}.
Such dusty star-burst galaxies in a protocluster core are predicted to merge into a single massive galaxy like a
Brightest Cluster Galaxy (BCG) as seen in the local universe \citep{rennehan20}.
Interestingly, \citet{rotermund21} found that a protocluster core at $z=4.3$ composed of many dusty star-burst
galaxies does not exhibit a high overdensity of dropout galaxies.
Although it is difficult to determine which galaxy population is the better tracer of environments or cosmic
structure, it is clear that steady galaxy evolution represented by dropout galaxies and stochastic phases like
experienced by star-bursting objects can occur in (locally) different environments.
The spatial segregation of galaxy populations makes it difficult to draw a conclusion on a protocluster's structure.

\section{Conclusions} \label{sec:con}

We have presented the spectroscopic confirmation of a protocluster at $z=3.699$ in the HSC-DUD layer.
This protocluster is composed of at least thirteen member galaxies.
As indicated by the photometric data \citep{toshikawa24}, spectroscopically-confirmed protocluster members tend to
be brighter in rest-frame UV than field counterparts.
One of these bright protocluster members is serendipitously identified to be an AGN though our protocluster search
does not depend on the presence of AGNs.
In addition, protocluster structure is investigated in terms of galaxies' separation from the centre of the
protocluster, $D_\mathrm{cen}$, and from their nearest neighbours, $D_\mathrm{nei}$.
The protocluster has a centrally-concentrated spatial distribution of member galaxies as indicated by a correlation
between $D_\mathrm{cen}$ and $D_\mathrm{nei}$.

We next compared galaxy properties ($M_\mathrm{UV}$ and Ly$\alpha$ $EW_0$) with the spatial parameters
$D_\mathrm{cen}$ and $D_\mathrm{nei}$.
The galaxy properties $M_\mathrm{UV}$ and Ly$\alpha$ $EW_0$ are more tightly related to $D_\mathrm{nei}$.
Since $M_\mathrm{UV}$ and Ly$\alpha$ $EW_0$ originate from strong radiation from massive stars, this suggests that
star-forming activity is more sensitive to smaller-scale environments rather than the position within the overall
protocluster.
In addition to the presence of an AGN, the environmental enhancement of star-forming activity results in an
overabundance of bright galaxies.

In order to search for other galaxy populations, that are not traced by the dropout technique, we have made use of
the public dataset of the COSMOS2020 and Super-deblended catalogues in the COSMOS field.
Massive, or mature, galaxies selected from the COSMOS2020 catalogue exhibit a higher overdensity near the centre of
the protocluster, implying that the protocluster has a higher galaxy density at the centre than observed via
Ly$\alpha$-emitting $g$-dropout galaxies.
In contrast, dusty star-burst galaxies selected from the Super-deblended catalogue reside in the outskirts of the
protocluster.
Although it is beyond the scope of this study to reveal the physical origin of this spatial segregation depending
on the galaxy population from star-burst to mature galaxies, our analysis demonstrates a complex of multiple galaxy
populations even at $z=3.7$.

This paper reports follow-up spectroscopy on a protocluster candidate among $N\sim100$ samples in the HSC-DUD
layer, as a pilot observation.
Even from this single case, we have obtained an insight into the relation between galaxy evolution and cluster
formation.
In the future, we will expand this campaign of follow-up spectroscopy to address the general trend and physical
origin of the diversity of cluster formation.

\section*{Acknowledgements}
JT and SW acknowledge support from STFC through grant ST/T000449/1.

\section*{Data Availability}
The data underlying this article will be shared on reasonable request to the corresponding author.

\bibliographystyle{mnras}
\bibliography{refs}

\bsp	
\label{lastpage}
\end{document}